# TeV Gamma Rays Expected from Supernova Remnants in Different Uniform Interstellar Media

E.G. Berezhko [a] and H.J. Völk [b]

[a] *Institute of Cosmophysical Research and Aeronomy, Lenin Ave. 31, 677891 Yakutsk, Russia*

[b] *Max-Planck-Institut für Kernphysik, Postfach 103980, D-69029 Heidelberg, Germany*

*Keywords:* Cosmic rays, supernova remnants, shock acceleration, $\gamma$-rays

**Abstract**

Calculations of the expected TeV $\gamma$-ray emission, produced by accelerated cosmic rays (CRs) in nuclear collisions, from supernova remnants evolving in a uniform interstellar medium (ISM) are presented. The aim is to study the sensitivity of $\gamma$-ray production to a physical parameter set. Apart from its general proportionality to $N_H$, it is shown that the $\gamma$-ray production essentially depends upon the ratio of the CR diffusion coefficient $\kappa$ to a critical value $\kappa_{crit} = 10(B_0/5\ \mu\text{G})(N_H/0.3\ \text{cm}^{-3})^{-1/3}\kappa_B$, where $B_0$ and $N_H$ are the magnetic field and hydrogen number density of the ISM, and $\kappa_B$ denotes the Bohm diffusion coefficient. If $\kappa$ is of the same order or lower than $\kappa_{crit}$, then the peak TeV $\gamma$-ray flux in the Sedov evolutionary phase, normalized to a distance of 1 kpc, is about $10^{-10}(N_H/0.3\ \text{cm}^{-3})$ photons cm$^{-2}$ s$^{-1}$. For a CR diffusion coefficient that is significantly larger than $\kappa_{crit}$, the CR cutoff energy is less than 10 TeV and the expected $\gamma$-ray flux at 1 TeV is considerably below the presently detectable level of $10^{-12}$ photons cm$^{-2}$ s$^{-1}$. The same is of course true for a SNR in the rarified, so-called hot ISM.

## 1 Introduction

A direct empirical test, whether or not the observed Galactic cosmic rays (CRs) are indeed produced in supernova remnants (SNRs) at least up to an energy of about $10^{14}$ eV (see e.g. [1-3] for reviews of CR acceleration theory) should be possible with observations of SNRs in high energy ($\epsilon_\gamma \sim 1$ TeV) $\gamma$-rays. If the dominant nuclear component of the CRs is strongly enhanced inside SNRs, then through hadronic collisions, leading to pion production and subsequent decay, $\gamma$-rays will be strongly produced.

Theoretical estimates of the $\pi^0$-decay $\gamma$-ray luminosity of SNRs [4–6] have led to the conclusion that the expected TeV $\gamma$-ray flux from nearby SNRs in high enough ambient densities should be just detectable by present instruments. The conclusions were based on a hydrodynamic approximation for the CR component and involved (very reasonable) assumptions about the CR energy spectra.

The spectra of $\pi^0$-decay $\gamma$-rays, produced by shock accelerated CRs in SNRs that expand into a the uniform interstellar medium (ISM), were studied in detail in a previous paper [7] (hereafter referred to as Paper I) in a kinetic approach (e.g. [8]).

The kinetic model prediction for the peak value of the expected $\gamma$-ray flux is not very different from that obtained in the simplified models [5], even though this difference is not unimportant. The main reason is that this peak value is mainly determined by the fraction of the explosion energy that is converted into CR energy, i.e. by the efficiency of CR acceleration, which is not strongly dependent on the model used.

There are more important differences in the time variation of the predicted $\gamma$-ray fluxes. Kinetic theory revealed a much more effective CR and therefore $\gamma$-ray production during the free expansion phase of the SNR, and a more rapid decrease of the $\gamma$-ray flux after reaching its peak value during the subsequent Sedov phase, due to the effect of different spatial distributions of the gas and the CRs inside the SNR (the so-called overlapping effect) that had not been taken into account in [5].

In Paper I we have considered only one particular set of possible physical parameters. In the present calculations we demonstrate how the expected TeV-energy $\gamma$-ray flux from SNRs depends upon the ISM density, the ejected mass and the CR diffusion coefficient.

## 2   Results and Discussion

We use here the full time-dependent, kinetic model for particle acceleration in SNRs which selfconsistently describes diffusive shock acceleration of CRs, taking account the nonlinear CR backreaction on the structure and dynamical evolution of the expanding spherical supernova (SN) shock [8]. CRs naturally originate from a suprathermal gas particle population at the shock front. This means that after shock heating the most energetic gas particles become involved, i.e. "injected", into the acceleration process.

Detailed investigations of CR acceleration and SNR evolution in the uniform ISM have demonstrated important features of this process [8]. For a wide range of possible injection rates the CR acceleration efficiency is very high and almost independent of the injection rate. This is why we use here the particular value of the injection parameter $\eta = 10^{-4}$, which denotes the fraction of gas particles involved in the acceleration. This value of $\eta$ provides an injection rate which is more than an order of magnitude lower compared with results of collisionless shock plasma simulations [9,10] and analytical injection theory [11], and with an injection rate that corresponds to the kinetic Monte Carlo-model (e.g. [12]) for purely parallel shocks. Our chosen value of $\eta$ effectively takes the influence of the shock obliquity into account: according to [12,13], already at angles $\theta \approx 45°$ between the upstream magnetic field and the shock normal the injection rate is about an order of magnitude smaller than in the purely parallel shock case.

As such the process of CR acceleration and associated $\gamma$-ray production in SNRs is detailed in Paper I. Here we present only calculations of the expected TeV $\gamma$-ray emission, measureable by the imaging Cherenkov technique. In order to illustrate the sensitivity of the $\gamma$-ray production to relevant physical parameters of the ISM, we present in Fig. 1 calculations of the expected integral $\gamma$-ray flux $F_\gamma(> 1 \text{ TeV})$, normalized to the distance $d = 1$ kpc, for eight different parameter sets. Four of them correspond to physical parameters typical for the case of a SN Ia: SN explosion energy $E_{sn} = 10^{51}$ erg, ejecta mass $M_{ej} = 1.4 M_\odot$, and the value $k = 7$ of the parameter $k$ which describes the ejecta velocity distribution; four other parameters correspond to the core collapse SN Ib /SN II cases: $E_{sn} = 10^{51}$ erg, $M_{ej} = 10\ M_\odot$, $k = 10$.

Note that an essential part of the volume of our Galaxy is occupied by a rarefied, so-called hot ISM phase, with a density that is about two orders of magnitude lower than another representative phase, the so-called warm ISM phase, with hydrogen number density $N_H = 0.3$ cm$^{-3}$. The kinetic theory results imply that CRs are effectively produced also by SNRs in the hot ISM. But even simple estimates show that in this case the expected $\pi^0$-decay $\gamma$-ray flux is far below practical detection possibilities, due to the extremely low ISM density (there might be detectable Inverse Compton emission, which is however not considered here). Therefore, in both cases above, the calculations were performed for an ISM hydrogen number density $N_H = 0.3$ cm$^{-3}$ and 30 cm$^{-3}$. The latter case can model a SNR evolving inside a rather dense cloud.

SN 1006, the remnant of a SN Ia, is the only SNR where TeV $\gamma$-ray emission was definitely claimed to have been detected up to now [14]. It exhibits also strong X-ray synchrotron emission with a spectrum that implies a maxi-mum electron energy $\epsilon_{max} \gtrsim 10$ TeV. According to the calculations in [15], this should also be the maximum energy of the proton component which implies that CR diffusion is about ten times less effective than Bohm diffusion. Therefore we shall also consider two different values of the CR diffusion coefficient. The first corresponds to the ordinary Bohm-type diffusion coefficient $\kappa = \kappa_B$ (e.g. Paper I); in the second case we use a ten times higher value, $\kappa = 10\kappa_B$. Note that different values of $\kappa$ influence only the value of the maximum CR energy $\epsilon_{max}$ (or maximum momentum $p_{max} = \epsilon_{max}/c$, where $c$ is the speed of light), reached during SNR evolution. In all cases we use the same standard value of the ISM magnetic field, $B_0 = 5$ $\mu$G.

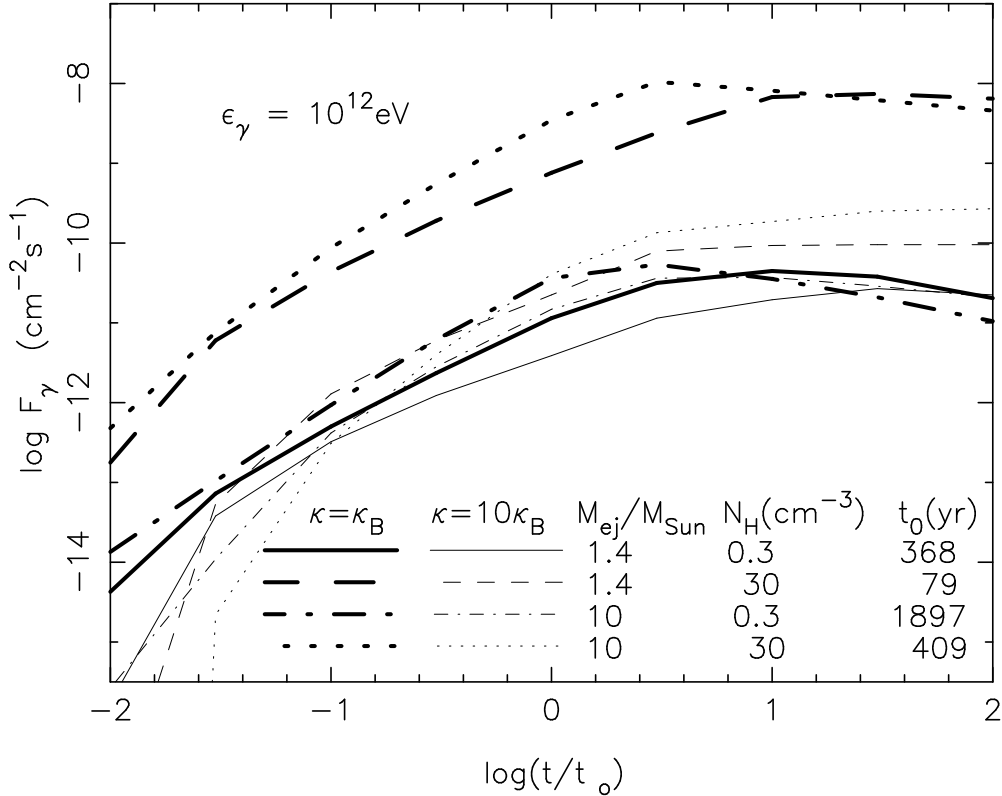

Fig. 1. Integral TeV γ-ray flux, normalized to the distance $d = 1$ kpc, as a function of time. Thick lines correspond to the Bohm diffusion coefficient $\kappa_B(p)$, thin lines correspond to $\kappa = 10\kappa_B$. Four different cases correspond to the following 3-parameter set: $M_{ej}/M_\odot$, $N_H/(1\ \text{cm}^{-3})$, $t_0/(1\ \text{yr})$): 1.4, 0.3, 368 (full lines); 1.4, 30, 79 (dashed lines); 10, 0.3, 1897 (dash-dotted lines); 10, 30, 409 (dotted lines).

To the extent that the CR energy $\epsilon \approx 10\epsilon_\gamma \approx 10$ TeV, responsible for γ-rays with energy $\epsilon_\gamma = 1$ TeV, is close to $\epsilon_{max}$, $\kappa$ essentially influences the expected γ-ray flux $F_\gamma(> 1\ \text{TeV})$: in the case $\epsilon_{max} < 10$ TeV $F_\gamma(> 1\ \text{TeV})$ should be essentially lower than at $\epsilon_{max} \gtrsim 10$ TeV. Therefore we present in Fig. 2 the calculated value of the maximum CR momentum $p_{max}$ as a function of time. By definition $p_{max}$ is the CR momentum, where the overall momentum spectrum $N(p, t)$ of CRs, accelerated up to time $t$, deviates from the dependence $p^{-2}$ by

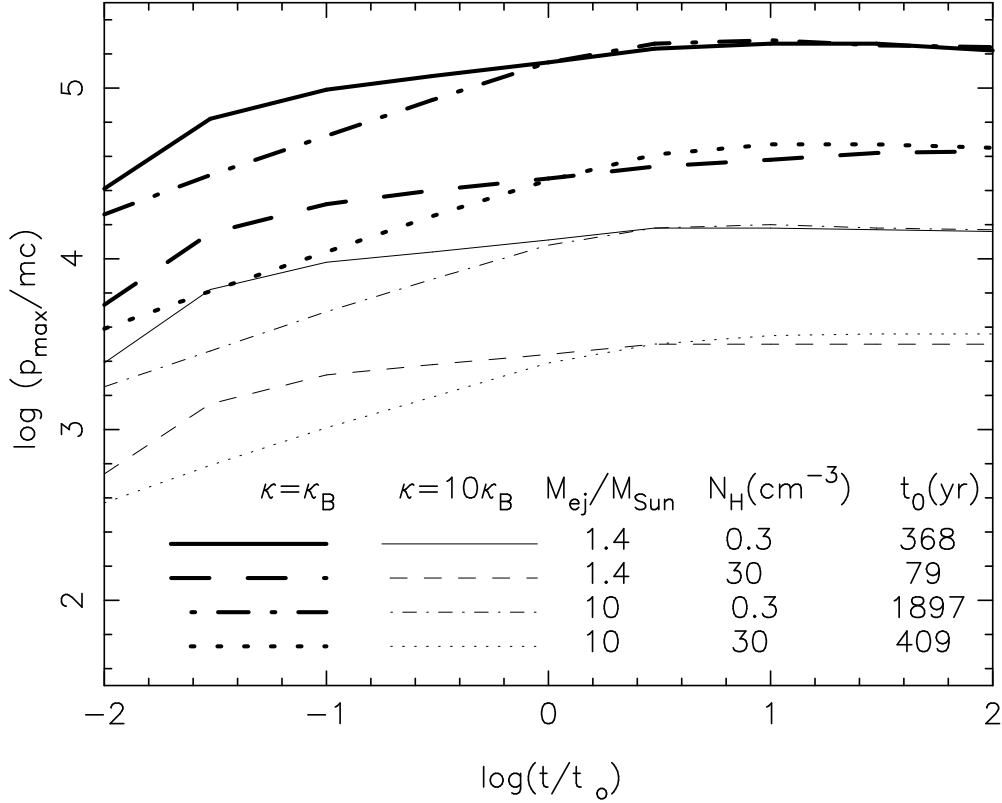

Fig. 2. The maximum CR momentum as a function of time for the same cases as in Fig. 1.

a factor of $e$:

$$\frac{N(p_{max},t)}{N(mc,t)}\left(\frac{p_{max}}{mc}\right)^2 = \frac{1}{e}. \qquad (1)$$

For the sake of convenience, in Figs. 1 and 2 time is measured in units of $t_0$, where $t_0$ is related to the sweep-up radius $R_0$ and the mean initial ejecta speed $V_0$ as follows:

$$t_0 = \frac{R_0}{V_0}, \qquad R_0 = \left(\frac{3M_{ej}}{4\pi\rho_0}\right)^{\frac{1}{3}}, \qquad V_0 = \sqrt{\frac{2E_{sn}}{M_{ej}}}. \qquad (2)$$

Here $\rho_0 = 1.4\,N_H m$ is the density of the ISM which contains 10% helium, and $m$ is the proton mass.

The maximum momentum $p_{max}$ is determined by geometrical factors [16]. The CR spectrum starts to deviate significantly from its power law form at momenta $p \sim p_{max}$, where the shock becomes too slow - and too small in radial extent - to fill the upstream region with CRs with a number density that is sufficient to provide a power law spectrum. The shock size $R_s$ and its speed $V_s$ are the most relevant parameters which influence the value of $p_{max}$. In the free expansion phase, $t < t_0$, $p_{max} \propto R_s V_s$ is an increasing function of time [16]. During the Sedov phase, $t \gtrsim t_0$, $p_{max}$ remains almost constant, because the shock weakens at this stage, the value of $R_s V_s \propto t^{-1/5}$ decreases with time, and the shock produces CR particles only up to the momentum $p_m \propto R_s V_s$ which is a decreasing function of time. During this phase the value of $p_{max}$ is determined by the CRs accelerated at the end of the free expansion phase. For $t \gg t_0$ the influence of the shock on these highest energy CRs (the so-called escaping particles) becomes almost negligible, and their energy remains nearly constant. The calculated results correspond very well to the expected dependence [16]:

$$p_{max} \propto E_{sn}^{1/2} M_{ej}^{-1/6} N_H^{-1/3}/(\kappa_B/\kappa). \qquad (3)$$

However this relation was derived in the test particle approximation and does not take into account the nonlinear effects which also influence $p_{max}$. As one can see from Fig. 2, the exact value of $p_{max}$ in the Sedov phase is slightly higher for a larger ejected mass $M_{ej}$, contrary to relation (3): the modified SN shock produces high energy CRs more efficiently and this increases their maximum momentum by a factor of a few compared with the case of an unmodified shock. For example, in the case $\kappa = \kappa_B$, $N_H = 0.3$ cm$^{-3}$, and $M_{ej} = 10 M_\odot$, the shock becomes significantly modified already at the end of the free expansion phase. The shock compression ratio is $\sigma = 5.5$ at $t = t_0$, and reaches its maximum value 5.8 at $t = 2t_0$. Therefore the maximum CR momentum reaches $p_{max} \approx 2 \times 10^5 mc$ soon after the beginning of the Sedov phase (see Fig. 2). It is slightly larger than the estimate $1.6 \times 10^5 mc$ predicted by test particle theory [16]. Due to the higher mean ejecta speed $V_0$ in the case $M_{ej} = 1.4 M_\odot$, the shock reaches peak modification significantly later: it is only slightly modified ($\sigma = 4.5$) at $t = t_0$ and the maximum shock compression ratio $\sigma = 5.8$ is reached at $t = 10 t_0$. At this stage, when the shock produces CRs most effectively, the product $R_s V_s$ has a lower value than at $t = t_0$. Therefore the maximum value $p_{max} = 1.8 \times 10^5 mc$ reached at $t = 10 t_0$ is slightly smaller than the test particle prediction $2.2 \times 10^5 mc$. The effect described gives in the Sedov phase a maximum CR momentum which is higher for larger ejecta mass, opposite to the test particle prediction (3), although the exact value of $p_{max}$ does not deviate essentially from the test particle prediction.

As can be seen from Fig. 2, for $N_H = 0.3$ cm$^{-3}$ the maximum CR momentum remains above the critical level $p_{max}(t) \gtrsim 10^4 mc$ at least for $t/t_0 \gtrsim 0.1$ even for a large diffusion coefficient $\kappa = 10 \kappa_B$. Therefore the value of $\kappa$ does not influence so much the expected flux $F_\gamma(> 1 \text{ TeV})$. In this case the predicted flux $F_\gamma$ is roughly proportional to the ISM density over the whole SNR evolution. The time profile of this integral $\gamma$-ray flux above 1 TeV is denoted by $F_\gamma(t/t_0)$ for short, and is roughly the same in all four cases. Note that the essential part of the dependence of $F_\gamma(t/t_0)$ on $M_{ej}$ and $N_H$ is already contained in the time unit $t_0$. The remaining part of the dependence on $M_{ej}$ is the dependence of the time $t_m$ (at which the peak value of the $\gamma$-ray flux is reached) upon $M_{ej}$. As one can see from Fig. 1 (again using $N_H = 0.3$ cm$^{-3}$), in the case $\kappa = \kappa_B$ $t_m/t_0 \approx 2.5$ for $M_{ej} = 10 M_\odot$, and $t_m/t_0 \approx 10$ for $M_{ej} = 1.4 M_\odot$. In the case $\kappa = 10\kappa_B$, $t_m/t_0 \approx 6.5$ for $M_{ej} = 10 M_\odot$, and $t_m/t_0 \approx 32$ for $M_{ej} = 1.4 M_\odot$. The reason is that one of the most relevant factors determining the value of the $\gamma$-ray flux is the CR energy content in the SNR. It reaches its peak value in the Sedov phase, when the shock speed $V_s$ drops to some critical value $V_m$, which is a function of the ISM parameters (see [7] for details). Since $t_0 \propto M_{ej}^{5/6}$, we have the relation $t_m/t_0 \propto M_{ej}^{-5/6}$. This corresponds satisfactorily to the results represented in Fig. 1.

Only in the case of a dense ISM, $N_H = 30$ cm$^{-3}$, does the increase of CR diffusion coefficient lead to a drop of the CR maximum momentum below the critical level $p_{max} < 10^4 mc$, that causes $F_\gamma(> 1$ TeV$)$ to drop significantly compared with its value in the case $\kappa = \kappa_B$. In this case TeV $\gamma$-rays are produced by CRs which belong to the exponential part of their spectrum $p \gtrsim p_{max}$. Therefore the $\gamma$-ray emissivity drops exponentially with increasing energy $\epsilon_\gamma$. Note that this significant influence of the CR diffusion coefficient on the expected $\gamma$-ray flux takes place only in the energy range $\epsilon_\gamma \gtrsim 1$ TeV for this density. For considerably lower energies $\epsilon_\gamma \ll 1$ TeV the $\gamma$-ray production is due to correspondingly lower-energy particles which belong to the power law part of CR spectrum, and thus the spectrum remains insensitive to the value of the CR diffusion coefficient. As can be seen from Fig. 1, the expected TeV $\gamma$-ray flux remains almost unchanged despite of an increase of the ISM density from $N_H = 0.3$ cm$^{-3}$ up to 30 cm$^{-3}$ in the case $\kappa = 10\kappa_B$. This implies that the effect of the $\kappa$-increase is about 100.

The highest energy CRs with momenta $p \gtrsim p_{max}$ become almost insensitive to the shock influence at the late Sedov phase $t \gg t_0$ [8]. They fill a volume almost uniformly whose size increases with time at this stage according to the diffusive law $R \propto \sqrt{\kappa t}$. Hence we have the situation when CRs with a number density $n(\epsilon) \propto R^{-3}$ interact with a progressively increasing amount of gas $M \propto R^3$, resulting in an almost constant TeV $\gamma$-ray flux $F_\gamma \propto Mn$ (see the curves, which correspond to $\kappa = 10\kappa_B$ and $N_H = 30$ cm$^{-3}$, in Fig. 1).

Our calculations show that the efficiency of TeV $\gamma$-ray production by shock accelerated CRs in a SNR is characterized by a critical value of the CR diffusion coefficient, that can be represented in the form

$$\kappa_{crit} = K \left(\frac{B_0}{5 \ \mu\text{G}}\right) \left(\frac{N_H}{0.3 \ \text{cm}^{-3}}\right)^{-1/3} \kappa_B, \qquad (4)$$

where the value $K \approx 10$ is expected to be only slightly dependent on the injection rate. For CR diffusion coefficients significantly larger than $\kappa_{crit}$, we expect maximum CR energies $\epsilon_{max} < 10$ TeV, which would lead to a sizeable decrease of the TeV $\gamma$-ray flux below a detectable level at all ISM densities considered.

The critical value of the CR diffusion coefficient exists for an arbitrary $\gamma$-ray energy $\epsilon_\gamma$. It is clear from the above that it is determined by the expression (4) with

$$\kappa_{crit} = \left(\frac{\epsilon_\gamma}{1\ \text{TeV}}\right)^{-1} \kappa_{crit}(1\ \text{TeV}). \tag{5}$$

This simply shows that the lower $\gamma$-ray energy $\epsilon_\gamma$ is, the wider is the range of the CR diffusion coefficient which allows efficient production of $\gamma$-rays with energy $\epsilon_\gamma$.

## 3  Summary

We have studied here how the background ISM density and the CR diffusion coefficient influence the expected $\pi^0$-decay $\gamma$-ray production in SNRs in the TeV range. Our calculations show that a peak TeV $\gamma$-ray flux $F_m$, normalized to a distance of $d = 1$ kpc, of about $10^{-10}$ photons cm$^{-2}$ s$^{-1}$ is reached at $t_m \simeq 3t_0$, for an ISM number density $N_H = 0.3$ cm$^{-3}$. In the Bohm limit for CR diffusion the ratio $F_m/N_H$ is independent of ISM density in the range $N_H = 0.3 \div 30$ cm$^{-3}$. If the $\gamma$-ray flux is written in the form $F_\gamma(t/t_0)$, then the ejected mass $M_{ej}$ influences only the peak time $t_m$, with $t_m/t_0 \propto M_{ej}^{5/6}$.

If CR scattering is ten times less efficient near the shock front, then the maximum CR energy decreases below the critical value 10 TeV which makes the $\gamma$-ray production almost independent of the ISM density, at a level that corresponds to $N_H = 0.3$ cm$^{-3}$ in the Bohm limiting case.

If the CR diffusion coefficient is significantly larger than $\kappa_{crit}$ (see eq.(4)), then the expected TeV $\gamma$-ray flux will drop considerably below the presently detectable level of $10^{-12}$ photons cm$^{-2}$ s$^{-1}$. Note that in this case SNRs would also be hardly considered as the sources of the Galactic CRs.

*Acknowledgments.* This work has been supported in part from Russian Foundation of Basic Research grant 97-02-16132. One of the authors (EGB) gratefully acknowledges the hospitality of the Max-Planck-Institut für Kernphysik where part of this work was carried out.